# Four ribbons of double-layer graphene suspending masses for NEMS applications

Xuge Fan[1,2,3]*, Chang He[1], Jie Ding[2]*, Sayedeh Shirin Afyouni Akbari[4] and Wendong Zhang[5,6]*

[1]Advanced Research Institute of Multidisciplinary Sciences, Beijing Institute of Technology, 100081 Beijing, China.

[2]School of Integrated Circuits and Electronics, Beijing Institute of Technology, 100081 Beijing, China.

[3]Center for Interdisciplinary Science of Optical Quantum and NEMS Integration, Beijing 100081, China.

[4]Advanced NEMS Group, École Polytechnique Fédérale de Lausanne (EPFL), 1015 Lausanne, Switzerland.

[5]State Key Laboratory of Dynamic Measurement Technology, North University of China, Taiyuan 030051, China.

[6]National Key Laboratory for Electronic Measurement Technology, School of Instrument and Electronics, North University of China, Taiyuan 030051, China.

*Email: xgfan@bit.edu.cn, jie.ding@bit.edu.cn, wdzhang@nuc.edu.cn

**Abstract**

Graphene ribbons with a suspended proof mass for nanomechanical systems have been rarely studied. Here, we report three types of nanomechanical devices consisting of graphene ribbons (two ribbons, four ribbons-cross and four ribbons-parallel) with suspended Si proof masses and studied their mechanical properties. The resonance frequencies and built-in stresses of three types of devices ranged from tens of kHz to hundreds of kHz, and from 82.61 MPa to 545.73 MPa, respectively, both of which decrease with the increase of the size of proof mass. The devices with four graphene ribbons featured higher resonance frequencies and spring constants, but lower built-in stresses than two ribbon devices under otherwise identical conditions. The Young's modulus and fracture strain of double-layer graphene were measured to be 0.34 TPa and 1.13% respectively, by using the experimental data and finite element analysis (FEA) simulations. Our studies would lay the foundation for understanding of mechanical properties of graphene ribbons with a suspended proof mass and their potential applications in nanoelectromechanical systems.

**Introduction**

Graphene is a kind of honeycomb two-dimensional crystal and has atomic thickness and excellent properties such as high stiffness, low mass, high conductivity and good flexibility [1–3]. Thanks to unique mechanical and electrical performance, graphene is a promising material for manufacturing nanomechanical devices, featuring decreased dimension, high sensitivity, fast response time, etc.[4–9]. The earliest application of graphene in nanomechanical systems (NEMS) were resonators in 2007 by suspending a single-layer graphene sheets over a $SiO_2$ trench and actuating it by using either electrical or optical modulation [10]. In later studies, NEMS resonators based on suspended graphene without suspended mass were widely reported for studying the material and structural properties of graphene and the device applications [6,11–21]. In recent years, structures of suspended graphene without a suspended mass have been used for various types of NEMS pressure sensors, microphones, loudspeakers, hall sensors, mass sensors, gas sensors, and bolometers [8,22,23]. Further, the structures based on both doubly-clamped graphene ribbons with a suspended proof mass and fully-clamped graphene membranes with a suspended proof mass that can be used for NEMS accelerometers, vibrometers, and nonlinear mechanics of four graphene ribbons with a suspended proof mass at resonance started to be studied [24–30].

However, the impact of geometrical sizes of different types of graphene ribbons with a suspended proof mass on resonance frequencies, spring constant, quality factors and built-in stresses have not been studied yet, which would impact their device applications. Further, the maximum force that four graphene ribbons with a suspended proof mass can withstand before rupture and facture strain of suspended graphene ribbons have not been studied. In addition, the Young's modulus of double-layer graphene that are used in the structures of suspended graphene with a suspended proof mass for NEMS applications has rarely been studied.

In this paper, we report different types of nanomechanical structures based on double-layer graphene ribbons with a suspended Si mass with different geometrical sizes, including two ribbons with a suspended Si mass, four crossed ribbons with a suspended Si mass and four parallel ribbons with a suspended Si mass, respectively. We characterized the dynamic and static mechanical properties of these different types of devices by using Laser Doppler Vibrometer (LDV) measurements and atomic force microscope (AFM) indentation experiments, combining FEA simulations. In particular, we compared and discussed the resonance frequencies, quality factors, spring constants, built-in stresses of different types of graphene ribbons with suspended proof masses. And we obtained the Young's modulus and fracture strain of double-layer graphene by utilizing the combination of experimental data and FEA simulations. To be specific, we used the same device (device 17) for LDV measurements, followed by AFM tip indentation experiments and consequent FEA simulations, which ensure the accurate extraction of Young's modulus of double-layer graphene. After obtaining the values of built-in stress and Young's modulus of double-layer graphene in device 17, we used both values to obtain the fracture strain of double-layer graphene by using AFM tip indentation data and FEA simulations. All these results would contribute to deeply understanding the mechanical properties of graphene and thereby be useful for their potential applications in NEMS and related devices.

## Materials and methods

### Fabrication

We designed and fabricated three types of graphene ribbons with a suspended Si mass, including two ribbons with a suspended Si mass, four crossed ribbons with a suspended Si mass and four parallel ribbons with a suspended Si mass, respectively. The schematic diagram of the designs of three types of graphene ribbons with a suspended Si proof mass are shown in Fig. **1 a-c**. For device fabrication, the thermally oxidized silicon-on-insulator (SOI) wafer was used as device substrate, in which the thermally oxidized layer was 1.4 μm thick, and the device layer, BOX layer and handle layer is 15 μm thick, 2 μm thick and 400 μm thick, respectively. The oxidized silicon device layer was etched by lithography, reactive ion etching (RIE) and deep reactive ion etching (DRIE), which formed trenches and defined the Si proof mass (Fig. **1 d**). After this, likewise, the thermally oxidized handle layer of the SOI wafer was etched by lithography with backside alignment, RIE and DRIE etching to form the cavities and thereby expose the BOX layer of SOI to air (Fig. **1 e**).

A PMMA-based wet transfer method was used to integrate the chemical vapour deposition (CVD) graphene (Graphenea, Spain) from copper substrate to the pre-fabricated SOI substrate, in which the double-layer graphene was obtained by vertically stacking two monolayer graphene sheets on top of each other. The graphene ribbons were etched into desired shapes by using optical lithography and the low-power $O_2$ plasma etching (Fig. 1). Finally, the sections of the BOX layer (2 µm thick $SiO_2$ layer) of SOI substrate were sacrificially removed by dry plasma etching followed by vapour hydrogen fluoride (HF) etching, and thereby the Si proof mass attached to the graphene ribbons were released (Fig. **1 g**).

Optical microscopy and SEM imaging were used to observe and characterize the morphology of the fabricated devices. SEM images of the three types of ribbons with a suspended Si proof mass are shown in Fig. **1 h** (we call it "two-ribbon device" for short), Fig. 1 I (we call it "four-ribbon-cross device" for short) and Fig. **1 j** (we call it "four-ribbon-parallel device" for short). The Si proof masses in all type of devices were quadratic with side lengths ranging from 5 to 100 µm and the thickness of 16.4 µm. The trench widths of the fabricated devices that defined the length of the graphene ribbons ranged from 2 to 4 µm. The graphene ribbon width was 5 µm in all devices.

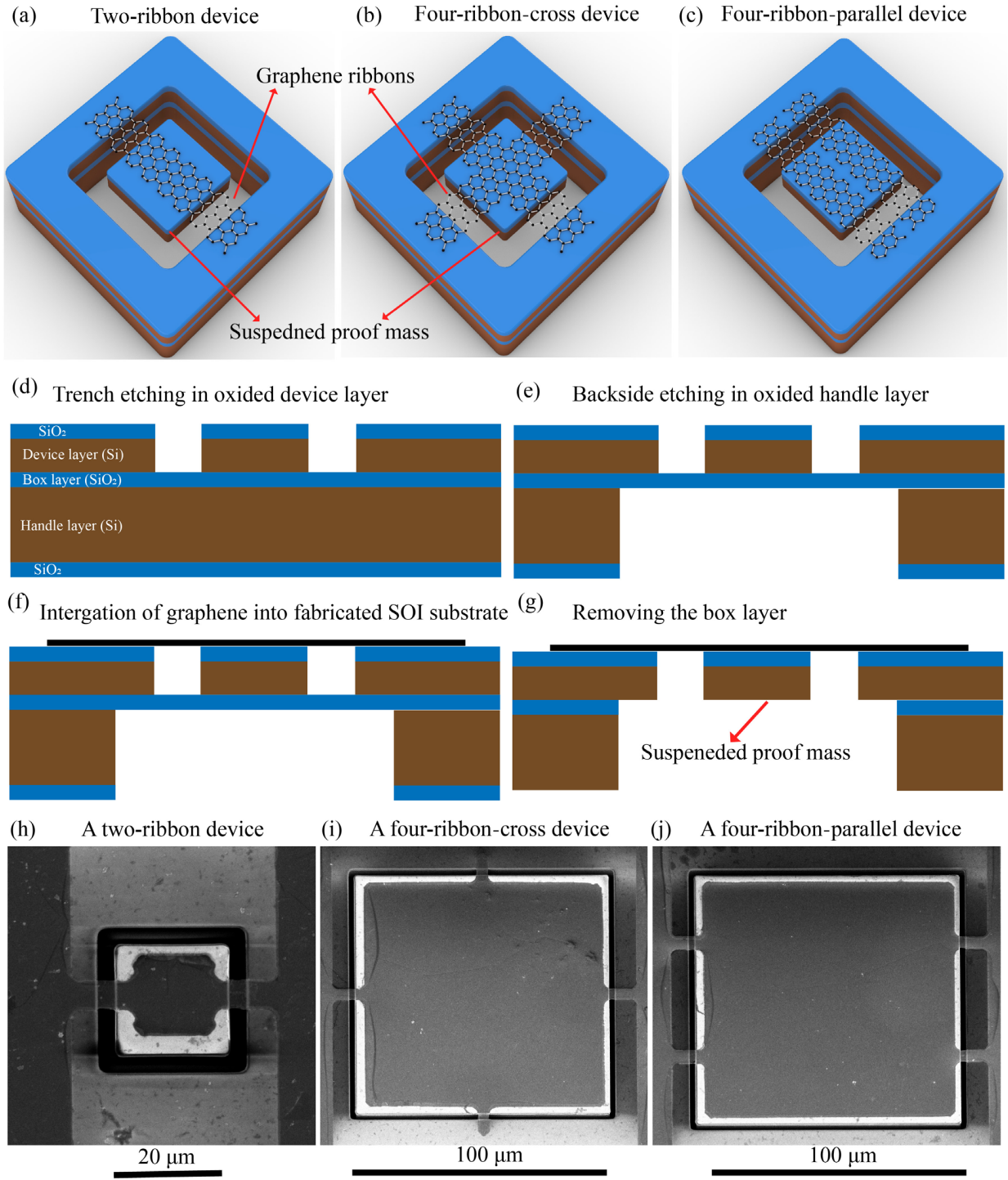

**Fig. 1 Different types of graphene ribbons with suspended Si proof masses. a-c** 3D schematics of the device design of two graphene ribbons with a suspended Si mass, four ribbons (cross) with a suspended Si mass, four ribbons (parallel) with a suspended mass. **d-g** Flow diagram of device

fabrication: d) the trenches were etched in the thermally oxidized silicon device layer of the SOI wafer thereby forming the Si mass; e) the section of handle layer of the SOI wafer was etched by DRIE to expose BOX layer to air; f) the double-layer graphene was integrated to the pre-fabricated SOI substrate by wet transfer method and etched into different types of ribbon structures by optical lithography and $O_2$ plasma etching; g) the oxidized Si proof mass was released by removing the BOX layer of the SOI substrate using RIE followed by vapour HF. **h-j** Top view of SEM images of three types of fabricated graphene ribbon devices: h) two ribbons with a suspended Si mass; i) four crossed ribbons with a suspended Si mass; and j) four parallel ribbons with a suspended Si mass .

## Results and discussion

### Dynamic mechanical characterization

To measure the dynamic mechanical properties of the spring-mass system of the fabricated graphene ribbon devices, we used a LDV (Polytec UHF-120) setup with a laser spot size on the order of 2.5 μm to measure the amplitudes of their thermomechanical noise in air (atmosphere pressure) at room temperature, which were then fitted to Lorentzian curves to estimate both resonance frequencies and quality factors [14,26]. It should be noted that all our measured ribbon devices were actuated by the laser-based thermal method. The details and dimensions of all measured devices are shown in Table **S1**. To analyse the elastic properties of the graphene ribbons, the effective spring constant (K) of the system that is associated with the resonance frequency and the quality factor can be expressed by[31]

$$K = m(2\pi f)^2 \tag{1}$$

where m is the mass, and f is the resonance frequency.

We characterized four two-ribbon devices (devices 1-4) with identical ribbon lengths (2 μm) and identical ribbon widths (5 μm), but different Si mass dimensions using LDV measurements as shown in Fig. **S1**. The extracted resonance frequencies of these devices were 65.5 kHz, 42.8 kHz, 39.7kHz and 22.7 kHz and the corresponding quality factors were 51.5, 60.6, 22.3 and 24.9 (Fig. **S1 a-d**). As expected from theory, the larger the attached mass of the two-ribbon device, the lower was the resulting resonance frequency (Fig. **S1 e**). There was no consistent trend regarding the quality factors of these devices (Fig. **S1 f**), indicating that the quality factors were dominated by other factors than the energy losses associated with the geometry of the attached masses [14,15]. The effective spring constants of the four two-ribbon devices (devices 1-4) were estimated to be 1.46 N/m, 1.11 N/m, 1.49 N/m and 0.71 N/m using equation 1, which were comparable to spring constants previously reported for graphene structures [10,24,32] and did not significantly depend on the dimensions of the proof mass.

To explore the impact of different graphene ribbon configurations in devices on their resonance frequency, quality factor and spring constant, we measured seven four-ribbon-cross devices (devices 5-11) and five four-ribbon-parallel devices (devices 12-16) with identical ribbon lengths (2 μm) and identical ribbon widths (5 μm), but different mass dimensions using LDV (Fig. **2** and Fig. **3**). The resonance frequencies of the seven four-ribbon-cross devices and the five four-ribbon-parallel devices decreased with increasing the size of the proof mass (Fig. **2 a-h** and Fig. **3 a-f**), but their quality factor did not show a consistent trend with respect to the proof mass (Fig. **2 i** and Fig. **3 g**). As the proof mass was quite large, the magnitude of the decrease of resonance frequency was weakened (Fig. **2 h**). Compared to the small proof mass, the large proof mass would generally result in decreased quality factor except few singularities probably due to more energy losses in the mechanical structure of four-ribbon-cross devices with large proof mass. The effective spring constant of the seven four-ribbon-cross devices were estimated to be 3.75 N/m, 3.71 N/m, 1.71 N/m, 2.31 N/m, 0.86 N/m, 0.59 N/m and 0.7 N/m. While the effective spring constant of the five four-ribbon-parallel devices were estimated to be 2.43 N/m, 2 N/m, 1.65 N/m, 0.55 N/m and 0.95 N/m. That is, the effective spring constant of four-ribbon-cross-devices and four-ribbon-parallel devices generally decreased with the increase of the size of the proof mass except few singularities. For the devices with identical ribbon length (2 μm), ribbon width (5 μm) and mass dimensions, the resonance frequency and spring constant of the four-ribbon-cross devices (devices 5-11) and four-ribbon-parallel devices (devices 12-16) were larger than those of the two-ribbon devices (devices 1-4).

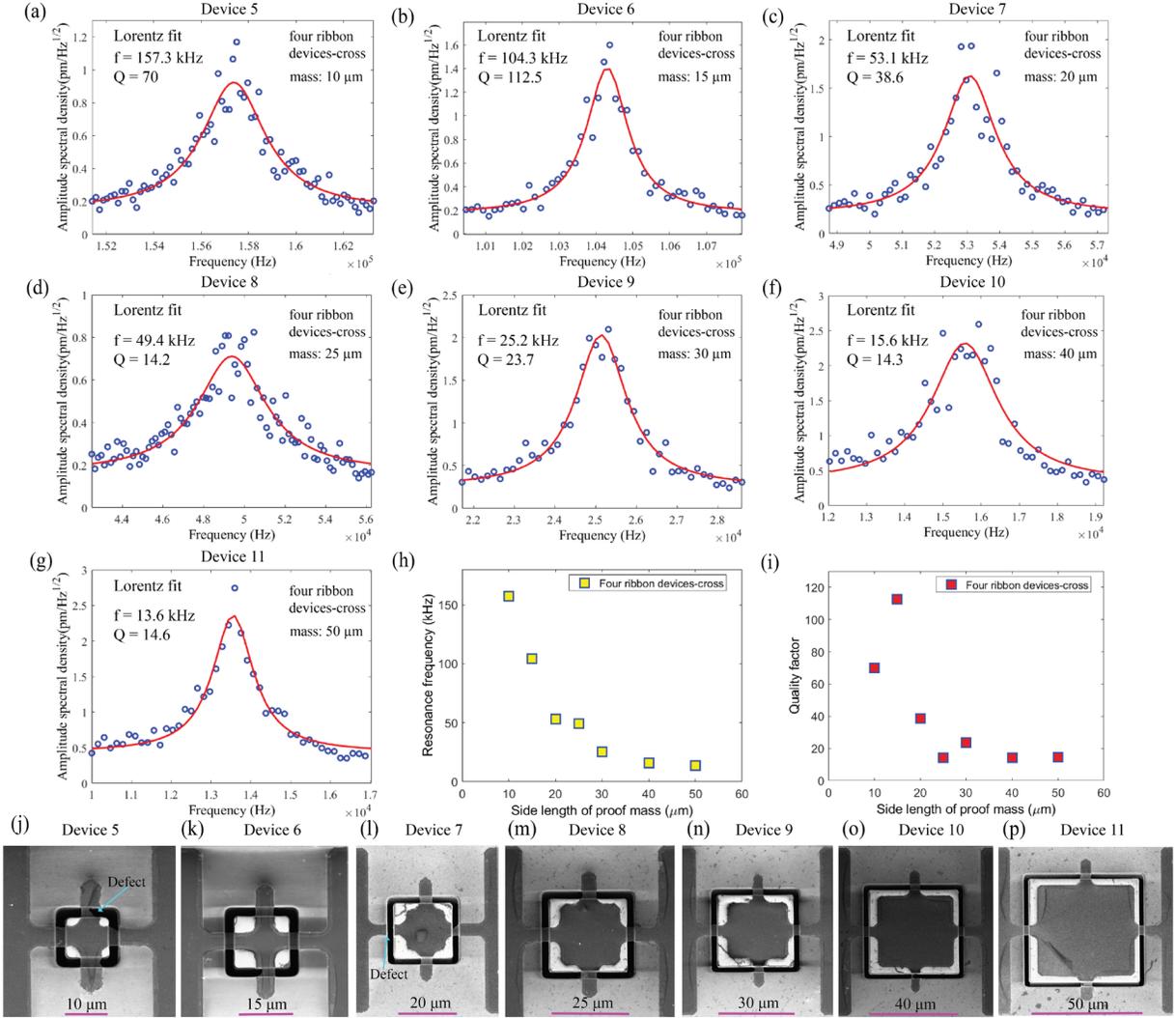

**Fig. 2 Dynamic mechanical characterization of four-ribbon-cross devices by measuring the amplitude of the thermomechanical noise in air using LDV. a-g** Thermomechanical noise peak of devices 5-11 using LDV, where the red solid lines in (a-g) were based on Lorentz fitting and extracted resonance frequencies and quality factors. The seven four-ribbon-cross devices have identical trench width (2 μm) and ribbon width (5 μm) but different proof mass dimensions (10 μm × 10 μm × 16.4 μm in (a); 15 μm × 15 μm × 16.4 μm in (b); 20 μm × 20 μm × 16.4 μm in (c); 25 μm × 25 μm × 16.4 μm in (d); 30 μm × 30 μm × 16.4 μm in (e); 40 μm × 40 μm × 16.4 μm in (f); 50 μm × 50 μm × 16.4 μm in (g)). **h** Resonance frequencies of devices 5-11 (157.3 kHz, 104.3

kHz, 53.1 kHz, 49.4 kHz, 25.2 kHz, 15.6 kHz and 13.6 kHz) versus the side length of the proof mass of devices 5-11. **i** Quality factor of devices 5-11 (70, 112.5, 38.6, 14.2, 23.7, 14.3 and 14.6) versus the side length of the proof mass of devices 5-11. **j-p** SEM images of devices 5-11 in (a-g), respectively. One graphene ribbon of device 5 (j) and one graphene ribbon of device 7 (l) had defects.

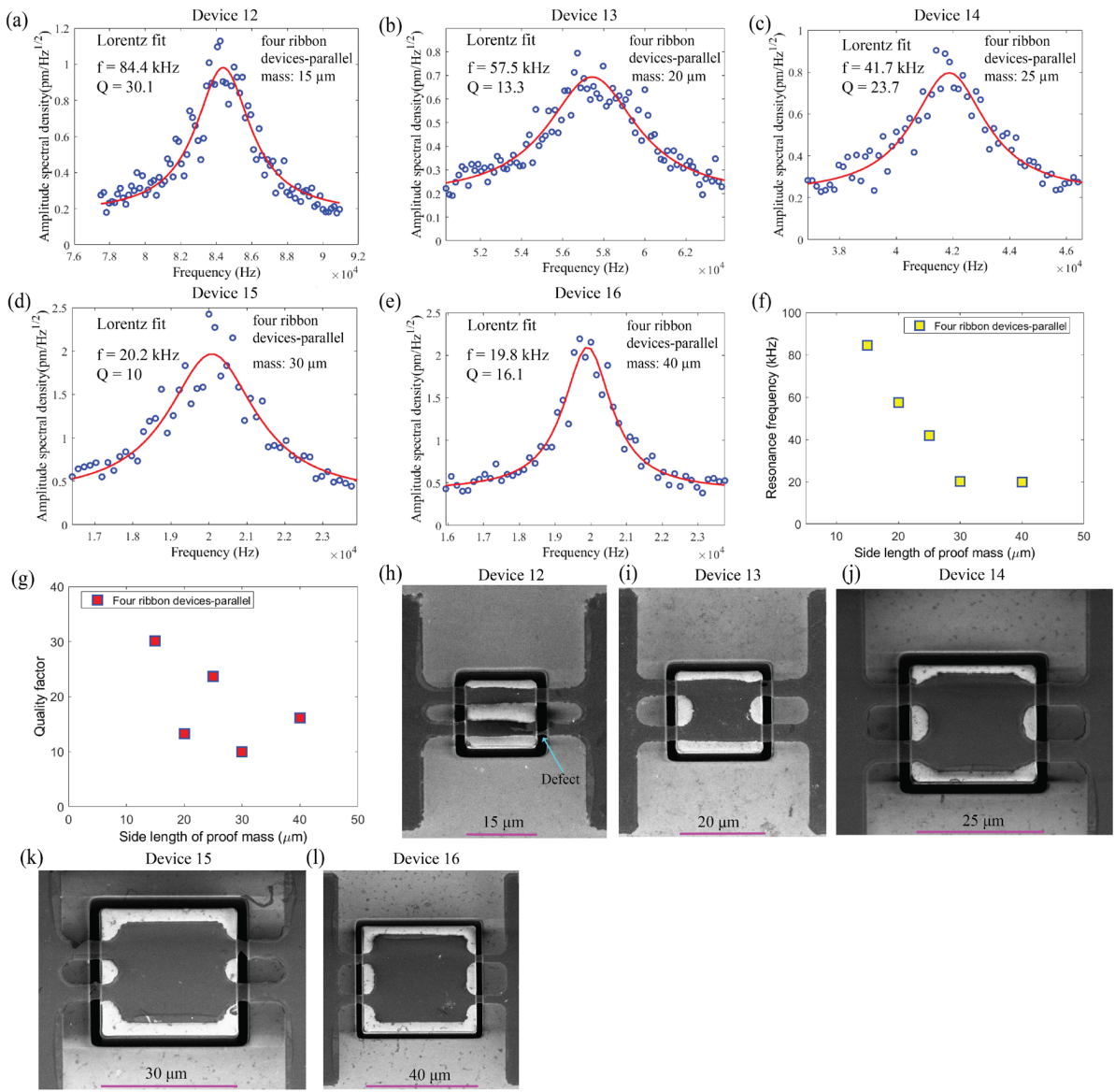

**Fig. 3 Dynamic mechanical characterization of four-ribbon-parallel devices by measuring the amplitude of the thermomechanical noise in air using LDV. a-e** Thermomechanical noise peak of devices 12-16 using LDV, where the red solid lines in (a-e) were based on Lorentz fitting and extracted resonance frequencies and quality factors. The five four-ribbon-parallel devices have identical trench width (2 µm) and ribbon width (5 µm) but different proof mass dimensions (15 µm × 15 µm × 16.4 µm in (A); 20 µm × 20 µm × 16.4 µm in (b); 25 µm × 25 µm × 16.4 µm in (c); 30 µm × 30 µm × 16.4 µm in (d); 40 µm × 40 µm × 16.4 µm in (e)). **f** Resonance frequencies of devices 12-16 (84.4 kHz, 57.5 kHz, 41.7 kHz, 20.2 kHz and 19.8 kHz) versus the side length of proof mass of devices 12-16. **g** Quality factor of devices 12-16 (30.1, 13.3, 23.7, 10 and 16.1) versus the side length of proof mass of devices 12-16. **h-l** SEM images of devices 5-11 in (a-e), respectively. One graphene ribbon of device 12 (h) had defects.

## Built-in stress of graphene ribbons

For application of graphene ribbons with a suspended proof mass in NEMS resonators and accelerometers, the built-in stress in the suspended graphene ribbons is an important characteristic of the structure. The built-in stress in suspended graphene ribbons typically is comparably large [8,24]. It would be of great significance to reduce the built-in stress in the graphene ribbons, because this could result in improved sensitivities of such devices [24,25]. At the same time, a decrease in the built-in stress in the graphene ribbons also means a decrease in the resonance frequency, which would limit the bandwidth of such devices, which means there exists a trade-off in these design parameters. Our analytical model shows that the built-in stresses of the two-ribbon devices were estimated to be hundreds of MPa (Text **S1**).

To further verify the built-in stress of the two-ribbon devices and explore the built-in stress of four-ribbon devices, we developed a finite element analysis (FEA) description by using the COMSOL and completed related simulation (Fig. **4 a**). As shown in Fig. **4 b**, by using a Young's modulus value of E = 0.22 TPa for double-layer graphene[24] and the measured resonance frequencies of devices, the FEA simulation results show that the built-in stresses in devices 1-4 are 412.29 MPa, 314.09 MPa, 418.29 MPa, 208.70 MPa, respectively, all of which are quite close to those values extracted by analytical model. Likewise, the FEA simulation results show that the built-in stresses in four-ribbon devices (devices 5-16) range from 82.61 MPa to 545.73MPa (Fig. **4 b**), which are on the same order of magnitudes with those of two-ribbon devices.

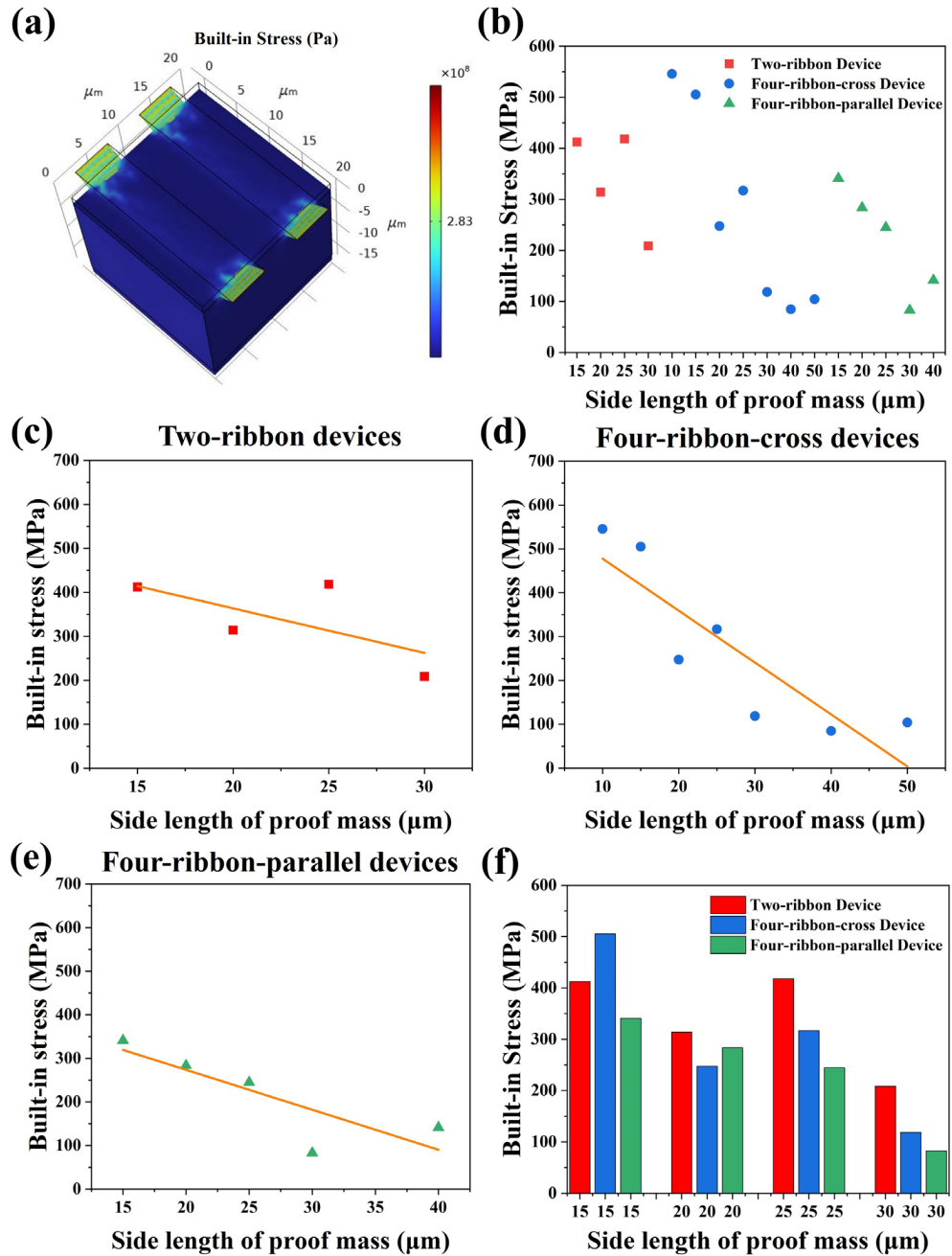

**Fig. 4 Built-in stresses in graphene ribbons of devices 1-16. a** 3D distribution of built-in stress in graphene ribbons of device 13 by COMSOL simulation. **b** Comparison of built-in stress values in graphene ribbons of devices 1-16, ranging from 82.61 MPa to 545.73 MPa. **c-e** The curve fittings by using data points of built-in stresses in graphene ribbons of two-ribbon devices (devices 1-4), four-ribbon-cross devices (devices 5-11), four-ribbon-parallel devices (devices 12-16),

respectively, indicating that the built-in stress generally decreases with the increase the size of the proof mass under otherwise identical conditions. The corresponding fitting equations are y=-10.13x+566.31 (c), y =-11.84x+596.19 (d), y= -9.16x+456.74 (e), where y is the built-in stress (unit: MPa) and x is the side length of the proof mass (unit: μm). **f** Comparison of built-in stress values in graphene ribbons of the different types of ribbon devices that have the same size of proof mass, trench width and ribbon length, indicating that the built-in stresses of two-ribbon devices are generally higher than four-ribbon-parallel devices.

To further study the impact of the magnitude of Young's modulus of graphene on the extraction of the built-in stress in devices, different Young's moduli of double-layer graphene were used for FEA simulation. The results demonstrate that the magnitude of Young's modulus has ignorable impact on the extraction of the built-in stress (Fig. **S2**).

To explore the impact of the geometrical design and the size of proof mass on the built-in stress of devices, the built-in stresses in devices that have the same type of devices, same trench width and ribbon length, but different sizes of the proof mass are compared (Fig. **4 c-e**). Except for very few singular data points probably due to the defects in some devices, the overall trend is that the built-in stress decreases with the increase of the proof mass under otherwise identical conditions, whether it is for two-ribbon devices (Fig. **4 c**) or four-ribbon-cross devices (Fig. **4 d**) or four-ribbon-parallel devices (Fig. **4 e**). In addition, the built-in stresses in devices that have the same trench width and ribbon length, the same size of the proof mass, but different type of devices are compared (Fig. **4 f**). It seems that the built-in stresses in two-ribbon devices are larger than those in four-ribbon-parallel devices.

**Static mechanical characterization**

To study the static mechanical properties and robustness of the four-ribbon devices, we performed force-displacement measurements by introducing indentation forces with an AFM tip at the centre of the suspended proof mass of a four-ribbon-cross device (device 17: ribbon length of 4 μm, proof mass size of 20 μm × 20 μm × 16.4 μm) (Fig. **5 a-c**). AFM indentation experiments were performed by using an AFM tool (Dimension Icon, Bruker) with a cantilever (Olympus AC240TM, calibrated spring constant: 5.303 N/m) and an AFM tip (tip radius = 15 nm). In these measurements we applied a defined AFM indentation force at the centre of the proof mass, whereafter we reduced the indentation force to 0 nN and then again applied an AFM indentation force that was higher than the previous AFM indentation force. And in experiments the AFM tip was placed at the same positions at the centre of the Si proof mass of measured devices. When the applied AFM indentation force in each consecutive loading/unloading cycle was gradually increased from 46.7 nN to 5368.5 nN, the maximum displacement of the Si mass in each loading cycle increased from 33.3 nm to 654 nm (Fig. 5 **b**). As the applied AFM indentation force is up to 5368.5 nN, the graphene ribbons did not break and the device was still intact. This indicates that these devices have strengths that is on the same order of magnitude as fully-clamped graphene membrane devices [25]. As the applied AFM indentation force is further increased to be 5800 nN, the graphene ribbons ruptured and the proof mass fell away from the graphene ribbons. This indicate that the maximum AFM indentation force that the graphene ribbons are able to withstand without rupture is around 5368 nN, which is consistent with those for fully-clamed graphene membranes and doubly-clamped graphene ribbons [26,33]. The displacement of graphene ribbons increased with the increase of the ribbon length, but decreased with the increase of the ribbon width [24,28,33]. The strain of graphene ribbons almost did not depend on the ribbon length, but decreased with the increase of the ribbon width [24,28,33].

In addition, dynamic mechanical characterization of device 17 by measuring the amplitude of thermomechanical noise in air using LDV was also performed before AFM indentation measurements (Fig. **5 d**). The extracted resonance frequency and quality factor were 55.2 kHz and 33.8, respectively. The effective spring constant of device 17 were estimated to be 1.84 N/m.

**Young's Modulus of double-layer graphene**

To accurately obtain the Young's modulus of double-layer graphene, we particularly used the same device (device 17) to do the LDV measurements followed by AFM tip indentation experiments. To be specific, the measured resonance frequency (55.2 kHz) of device 17 was used to extract the built-in stress of graphene ribbons in device 17. And the extracted built-in stress in device 17 was used together with the AFM tip indentation force-displacement measurement data of device 17 for the FEA simulations and fitting, which ultimately resulted in accurate extraction of Young's modulus of double-layer graphene. The results show that the Young's modulus of double-layer graphene in device 17 is around 0.34 TPa, with corresponding built-in stress of 531.33 MPa (Fig. **5 e**). For comparison, as the Young's modulus is set to be 1TPa and 0.22 TPa, respectively, the corresponding data points of force-displacement based on FEA simulation are far away from those measured based on AFM tip indentation. The Young's modulus of 0.34 TPa we obtained is much lower than the commonly  reported value of 1TPa for monolayer graphene [1,34]. This indicates that the Young's modulus of double-layer graphene is lower than monolayer graphene, which can be probably ascribed to the grain boundaries and ripples of the CVD graphene[35], different types and density of defects[36], and energy dissipation between graphene layers due to interlayer sliding[37,38].

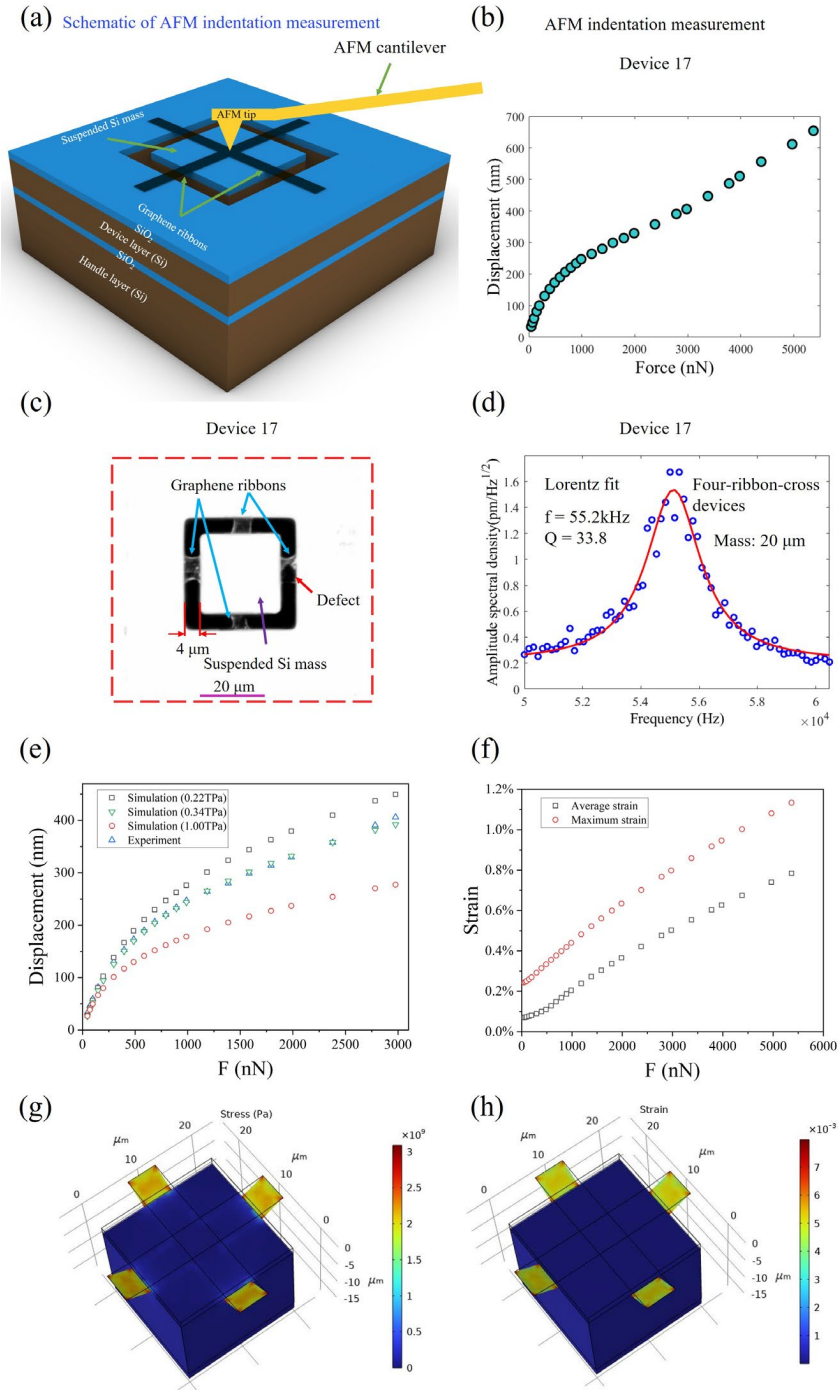

**Fig. 5 Force-displacement measurements of a four-ribbon-cross device with a suspended Si proof mass via AFM indentation experiments and the Young's modulus of double-layer graphene and the strain of the graphene ribbons. a** Schematic diagram of force-displacement measurements using AFM indentation at the centre of the suspended Si mass. **b** Force-

displacement measurement of a four-ribbon-cross device (device 17) with ribbon length of 4 μm and a proof mass of 20 μm × 20 μm × 16.4 μm. **c** High-contrast microscopy image of a four-ribbon-cross device (device 17) that was measured in (b). One graphene ribbon of device 17 had defects. **d** Thermomechanical noise peak of device 17 using LDV, where the red solid line was based on Lorentz fitting with extracted resonance frequency and quality factor. **e** The comparison of measured force-displacement data points of device 17 that was obtained by AFM tip indentation experiment and those of device 17 that was simulated by COMSOL with different Young's moduli of graphene (0.22 TPa, 0.34 TPa, and 1TPa). The results show that as the Young's modulus of graphene is 0.34 TPa with corresponding built-in stress of 531.33 MPa, the simulated data points is most close to the measured data points. Therefore, the Young's modulus of double-layer graphene is around 0.34 TPa. **f** The strain values of graphene ribbons of device 17 versus the applied force, including the average strain and maximum strain. **g** 3D distribution of stress in graphene ribbons of device 17 as the applied force is 2977 nN. **h** 3D distribution of strain in graphene ribbons of device 17 as the applied force is 2977 nN.

**Strain in graphene ribbons**

To study the strain of graphene ribbons, the average strain and maximum strain in the graphene ribbons of device 17 were obtained based on FEA simulation, by utilizing the Young's modulus of 0.34 TPa and built-in stress of 531.33 MPa we obtained from the same device (device 17). As shown in Fig. **5 f**, the simulation results show that the average strain is up to 0.784% while the maximum strain is up to 1.13%, as the applied force is set to be 5368.5 nN that is close to the maximum force that the graphene ribbons are able to withstand without rupture. Therefore, the fracture strain of graphene ribbons in device 17 can be estimated to be around 1.13%. The average strain of 0.78% in graphene ribbons of device 17 is at least one order of magnitude larger than the highest values of strain of graphene ribbons in other reports [32,39–41]. Further, the average strain of 0.78% in graphene ribbons of device 17 is also comparable with those in fully-clamped graphene membranes [1,26,34,42].

To clearly illustrate the stress and strain in graphene ribbons of device 17 during the process of AFM tip indentation experiments, Fig. **5 g** and **h** show the 3D distribution of stress and strain in graphene ribbons respectively, under the condition of the applied force of 2977 nN. The maximum stress and strain focus on the area of graphene ribbons that is close to the edges of the Si proof mass or the trench edges.

**Conclusion**

We have reported three types of graphene ribbons with suspended Si proof masses including two-ribbon devices, four-ribbon-cross devices and four-ribbon-parallel devices, which can be potentially used as NEMS transducers. We measured, compared and analysed the resonance frequencies, quality factors and spring constant of all three types of graphene ribbon devices. Based on measurement data and FEA simulation, we also obtained the built-in stresses of all three types of graphene ribbon devices, with typical values of hundreds of MPa, and found that the built-in stress generally decreases with the increase of the size of the proof mass under otherwise identical conditions. Also, we found that four-ribbon device design generally yield lower built-in stress than the two-ribbon device design. Further, we found that the four-ribbon device is able to withstand AFM indentation force of up to 5368.5 nN before the device was ruptured. We accurately obtained the Young's modulus of double-layer graphene of 0.34 TPa and the fracture strain of graphene ribbons of 1.13%. Compared to two-ribbon devices, the four-ribbon devices would potentially provide larger bandwidth, better mechanical stability, lower built-in stress, longer lifetime when they are designed as transducers for NEMS and device applications. These studies on different types of graphene ribbons with suspended Si proof masses would lay the foundation for understanding the properties of graphene and their potential applications in NEMS.

## Acknowledgements

This work was supported by the National Natural Science Foundation of China (Grant No. 62171037 and 62088101), 173 Technical Field Fund (2023-JCJQ-JJ-0971), Beijing Natural Science Foundation (4232076), National Key Research and Development Program of China (2022YFB3204600), National Science Fund for Excellent Young Scholars (Overseas), Beijing Institute of Technology Teli Young Fellow Program (2021TLQT012), Beijing Institute of Technology Science and Technology Innovation Plan (2022CX11019), the FLAG-ERA project 2DNEMS funded by the Swedish Research Council (VR) (2019-03412), and the Swiss National Science Foundation (project PP00P2_170590 and CRSII5_189967). The authors thank F. Niklaus and M. C. Lemme for discussion with device processing, and G. Villanueva for his help with LDV measurements.

## Conflict of Interest

The authors declare no competing interests.

## Supplementary Information

The online version of this article contains Supplementary Material available at xxx.

**Ethics approval and consent to participate**

Not applicable